\input harvmac
\input psfig
\input epsf
\noblackbox
\newcount\figno
 \figno=0
 \def\fig#1#2#3{
 \par\begingroup\parindent=0pt\leftskip=1cm\rightskip=1cm\parindent=0pt
 \baselineskip=11pt
 \global\advance\figno by 1
 \midinsert
 \epsfxsize=#3
 \centerline{\epsfbox{#2}}
 \vskip 12pt
 {\bf Fig.\ \the\figno: } #1\par
 \endinsert\endgroup\par
 }
 \def\figlabel#1{\xdef#1{\the\figno}}
 \def\encadremath#1{\vbox{\hrule\hbox{\vrule\kern8pt\vbox{\kern8pt
 \hbox{$\displaystyle #1$}\kern8pt}
 \kern8pt\vrule}\hrule}}
 %
 %


 \font\cmss=cmss10
 \font\cmsss=cmss10 at 7pt
 \def\rlx{\relax\leavevmode}
 \def\inbar{\vrule height1.5ex width.4pt depth0pt}
 \def\IC{\relax\,\hbox{$\inbar\kern-.3em{\rm C}$}}
 \def\IN{\relax{\rm I\kern-.18em N}}
 \def\IP{\relax{\rm I\kern-.18em P}}
 \def\ZZ{\rlx\leavevmode\ifmmode\mathchoice{\hbox{\cmss Z\kern-.4em Z}}
  {\hbox{\cmss Z\kern-.4em Z}}{\lower.9pt\hbox{\cmsss Z\kern-.36em Z}}
  {\lower1.2pt\hbox{\cmsss Z\kern-.36em Z}}\else{\cmss Z\kern-.4em
  Z}\fi}
 \def\IZ{\relax\ifmmode\mathchoice
 {\hbox{\cmss Z\kern-.4em Z}}{\hbox{\cmss Z\kern-.4em Z}}
 {\lower.9pt\hbox{\cmsss Z\kern-.4em Z}}
 {\lower1.2pt\hbox{\cmsss Z\kern-.4em Z}}\else{\cmss Z\kern-.4em
 Z}\fi}
 \def\IZ{\relax\ifmmode\mathchoice
 {\hbox{\cmss Z\kern-.4em Z}}{\hbox{\cmss Z\kern-.4em Z}}
 {\lower.9pt\hbox{\cmsss Z\kern-.4em Z}}
 {\lower1.2pt\hbox{\cmsss Z\kern-.4em Z}}\else{\cmss Z\kern-.4em Z}\fi}

 \def\narrowplus{\kern -.04truein + \kern -.03truein}
 \def\narrowminus{- \kern -.04truein}
 \def\narrowminussub{\kern -.02truein - \kern -.01truein}

 \def\frac#1#2{{#1\over #2}}

 \def\IZ{\relax\ifmmode\mathchoice
 {\hbox{\cmss Z\kern-.4em Z}}{\hbox{\cmss Z\kern-.4em Z}}
 {\lower.9pt\hbox{\cmsss Z\kern-.4em Z}}
 {\lower1.2pt\hbox{\cmsss Z\kern-.4em Z}}\else{\cmss Z\kern-.4em Z}\fi}
 \def\IB{\relax{\rm I\kern-.18em B}}
 \def\IC{{\relax\hbox{$\inbar\kern-.3em{\rm C}$}}}
 \def\Ic{{\relax\hbox{$\inbar\kern-.22em{\rm c}$}}}
 \def\ID{\relax{\rm I\kern-.18em D}}
 \def\IE{\relax{\rm I\kern-.18em E}}
 \def\IF{\relax{\rm I\kern-.18em F}}
 \def\IG{\relax\hbox{$\inbar\kern-.3em{\rm G}$}}
 \def\IGa{\relax\hbox{${\rm I}\kern-.18em\Gamma$}}
 \def\IH{\relax{\rm I\kern-.18em H}}
 \def\II{\relax{\rm I\kern-.18em I}}
 \def\IK{\relax{\rm I\kern-.18em K}}
 \def\IP{\relax{\rm I\kern-.18em P}}

 \font\cmss=cmss10 \font\cmsss=cmss10 at 7pt
 \def\IR{\relax{\rm I\kern-.18em R}}

 %

 %
 %
 \def\eqnn#1{\xdef #1{(\secsym\the\meqno)}\writedef{#1\leftbracket#1}%
 \global\advance\meqno by1\wrlabeL#1}
 \def\eqna#1{\xdef #1##1{\hbox{$(\secsym\the\meqno##1)$}}
 \writedef{#1\numbersign1\leftbracket#1{\numbersign1}}%
 \global\advance\meqno by1\wrlabeL{#1$\{\}$}}
 \def\eqn#1#2{\xdef #1{(\secsym\the\meqno)}\writedef{#1\leftbracket#1}%
 \global\advance\meqno by1$$#2\eqno#1\eqlabeL#1$$}

 \lref\author{Name}
\lref\berg{E.~Bergshoeff,
 D.~S.~Berman, J.~P.~van der Schaar and P.~Sundell, ``A
 noncommutative M-theory five-brane,'' hep-th/0005026.}

\Title
 {\vbox{
 \baselineskip12pt
 \hbox{HUTP-00/A047}
 \hbox{hep-th/0012041}\hbox{}\hbox{}
}}
 {\vbox{
 \centerline{Mirror Symmetry, D-branes}
 \vglue .5cm
 \centerline{and}
 \vglue .5cm
 \centerline{Counting Holomorphic Discs}
 }}
 \centerline{ Mina ${\rm Aganagic}$ and
Cumrun ${\rm Vafa}$}
 \bigskip\centerline{ Jefferson Physical Laboratory}
 \centerline{Harvard University}
\centerline{Cambridge, MA 02138, USA}
 \smallskip
 \vskip .3in \centerline{\bf Abstract}
{We consider a class of special Lagrangian subspaces of
Calabi-Yau manifolds and identify their mirrors, 
using the recent derivation of mirror symmetry, as certain
 holomorphic varieties of the mirror geometry.
This transforms the counting of holomorphic disc instantons
ending on the Lagrangian submanifold
to the classical Abel-Jacobi map on the mirror.  We recover
some results already anticipated as well as obtain some highly non-trivial
new predictions.
}
 \smallskip 
\Date{December 2000}
\newsec{Introduction}
Calabi-Yau geometry has been the source of many interesting
physical insights in string theory.  A key role is played by
mirror symmetry which relates questions involving the Kahler
geometry of the Calabi-Yau to complex geometry of a mirror
Calabi-Yau (or more generally complex parameters characterizing a
mirror description of the $N=2$ worldsheet theory).  A simple proof of
mirror symmetry has appeared in \ref\hova{K.~Hori and C.~Vafa,
``Mirror symmetry,''hep-th/0002222.}\ based on  enlarging
the gauge system of the linear sigma model \ref\witls{E.~Witten,
``Phases of N = 2 theories in two dimensions,''Nucl.\ Phys.\ 
{\bf B403}, 159 (1993) [hep-th/9301042].}\ 
and applying $T$-duality to the charged fields of the theory. 
It is thus natural to ask how this acts on the
D-branes. It is expected that even and odd branes of the two
geometries are exchanged under the mirror symmetry.  
This maps Lagrangian submanifolds (which are half the
dimension of the Calabi-Yau) on one side to the
complex submanifolds of the mirror geometry. Aspects of this
action were studied for certain massive sigma models in
\ref\hiv{K.~Hori, A.~Iqbal and C.~Vafa,``D-branes and mirror symmetry,''
hep-th/0005247.}.  The aim of this paper is to extend
mirror symmetry to certain special Lagrangian submanifolds of Calabi-Yau and
its mirror complex geometry. As a by-product we are able to
count the holomorphic discs ending on the Lagrangian submanifolds
using the Abel-Jacobi map of the mirror manifold.

The organization of this paper is as follows: In section 2 we
discuss aspects of toric geometry with emphasis on certain special
Lagrangian submanifolds associated to it. These constructions have
already appeared in the mathematics literature
\ref\hl{F.R.~Harvey and H.B.~Lawson,''Calibrated Geometries'',
Acta Math. (1982), 47-157.}\ref\jo{D.~Joyce,
``On counting special Lagrangian homology 3-spheres,''
hep-th/9907013.}\ and they are very natural from the
viewpoint of linear sigma models.  In section 3 we discuss mirror
symmetry, as derived in \hova , and apply it to the Lagrangian
submanifolds discussed in section 2 to obtain holomorphic
submanifolds of the mirror geometry.  In section 4 we show how the
holomorphic disc amplitudes of the A-model in certain cases are
related to Abel-Jacobi map of the mirror geometry.  We use this
result in section 5 to compute some holomorphic disc instanton
corrections.  In particular we confirm the result of
\ref\ov{H.~Ooguri and C.~Vafa,``Knot invariants and topological strings,''
Nucl.\ Phys.\  {\bf B577}, 419 (2000)[hep-th/9912123].}\ 
which predicts a  universal $1/n^2$
multi-covering formula for disc instantons. We also find highly
non-trivial predictions for the number of
holomorphic disc instantons in various situations
which pass the integrality check of \ov .

\newsec{Toric Geometry and Special Lagrangian Submanifolds}

We begin this section by briefly reviewing certain aspects of
toric geometry. Let $X={\bf C}^n$ be parameterized by
$x^1,\ldots,x^n$, and endowed with flat Kahler form $\omega =
i\sum_i dx^i \wedge d {\bar{x}}^i$. We can also view $\omega$ as
\eqn\kahl{\omega =\sum_i d|x^i|^2 \wedge d\theta^i}
where $\theta^i$ denotes the angular variable in the $x^i$ plane.

Consider a Lagrangian
submanifold $L=R^n$ of ${\bf C}^n$ corresponding to fixed
$\theta^i$. This $n$-dimensional real space is parametrized by
$|x^i|^2$ and the fact that it is Lagrangian follows trivially as
$\theta^i$ are constants, so  $\omega$ vanishes on it.  
Of course this description is valid as long as we are
away from loci where any $x^i=0$.  Note that ${\bf C}^n$ can be
viewed as a $T^n$ torus fibration over $L$, where the fibration
degenerates at the boundaries of $L$ (where any $|x^i|^2=0$).
This is the basic setup of toric geometry.

We can now describe other Lagrangian submanifolds of ${\bf C}^n$.  
Consider any submanifold  $D^r\in L$ of
dimension $r\leq n$.  For each point $p \in D^r$ consider the
$r$ dimensional
tangent space $T_p(D^r) \hookrightarrow T_p(L)$.  
This defines an $n-r$ dimensional subspace of the
fiber $T^n$ over that point, orthogonal with respect to $\omega$ 
to the tangent directions to $D^r$.  
If the slope of the subpace $D^r$ is rational
then the corresponding $n-r$ dimensional subpace of $T^n$
is a torus $T^{n-r}\subset T^n$ over $p$.  Let us
assume that $D^r$ has rational slope at all points--this
effectively reduces one to rational linear subspaces of $L$.  In this way
we obtain a Lagrangian submanifold associated to each such
subspace.  We can characterize a linear rational subspace of
$L$ by $n-r$ sets of $n$-tuple integers $q_i^\alpha$, where
$i=1,...,n$ and $\alpha =1,...,n-r$ such that
\eqn\lagbase{\sum_i q_i^{\alpha} |x^i|^2=c^{\alpha}}
where $c^{\alpha}$ are constants (not necessarily integers). One
can also write these in terms of $r$ vectors $v_\beta$ as
$$|x^i|^2=v^i_\beta s^\beta +d^i$$
where $\beta$ runs from $1,...,r$, $d^i$ are constants and
$$q^{\alpha}\cdot v_\beta =\sum_{i=1}^nq^{\alpha}_iv^i_\beta =0.$$
Note that the constraints on the $\theta^i$ are
\eqn\lagfiber{\sum_i v^i_\beta \theta^i =0}
and equivalently
$$\theta^i =q_i^{\alpha} \phi_{\alpha}.$$

Below, we will be interested in a subset of such Lagrangian
submanifolds known as {\it special} Lagrangian submanifolds, which
satisfy the property that for each $\alpha$
\eqn\conspla{\sum_{i=1}^{n} q_i^{\alpha}=0.}

So far we have ignored the discussion of boundaries of $L$ and
the other Lagrangian submanifolds, and 
whether the above constructions can be extended to true
Lagrangian submanifold without boundary. $L$ itself is not Lagrangian but
it will be if we take $2^n$ fold cover of it (by choosing,
for each $i$, both the $\theta^i=0$ section and $\theta^i=
\pi$ section of ${\bf C}^n$) and it will correspond to the real
subspace of ${\bf C}^n$.  
 
Similar statement holds for subspaces
$D^r\subset L$ with boundaries at $x^i=0$ for some of the $i$.  
But also sometimes it is not necessary to do this doubling.
Consider for example ${\bf C}^2$ in which case $L$ can
be identified with the positive quadrant of the 2 dimensional
plane. Consider the Lagrangian submanifold $D\subset L$ given by
$$(q_1,q_2)=(1,-1).$$
This corresponds to the subspace
$$|x^1|^2-|x^2|^2=c, \qquad \theta^1+\theta^2=0.$$
For generic $c>0$ where $|x^2|^2=0$ this meets the boundary of $L$
at $|x^1|^2=c$ and unless we double the geometry, $D$ will give
rise to a Lagrangian submanifold with boundary.
However if we consider the particular case where $c=0$ (see figure 1) 
then we do not need to double the geometry and $D$
corresponds to a Lagrangian submanifold without any boundaries. It
in fact corresponds to $x^1={\overline x}^2$.
\bigskip
\centerline{\epsfxsize 3.truein\epsfbox{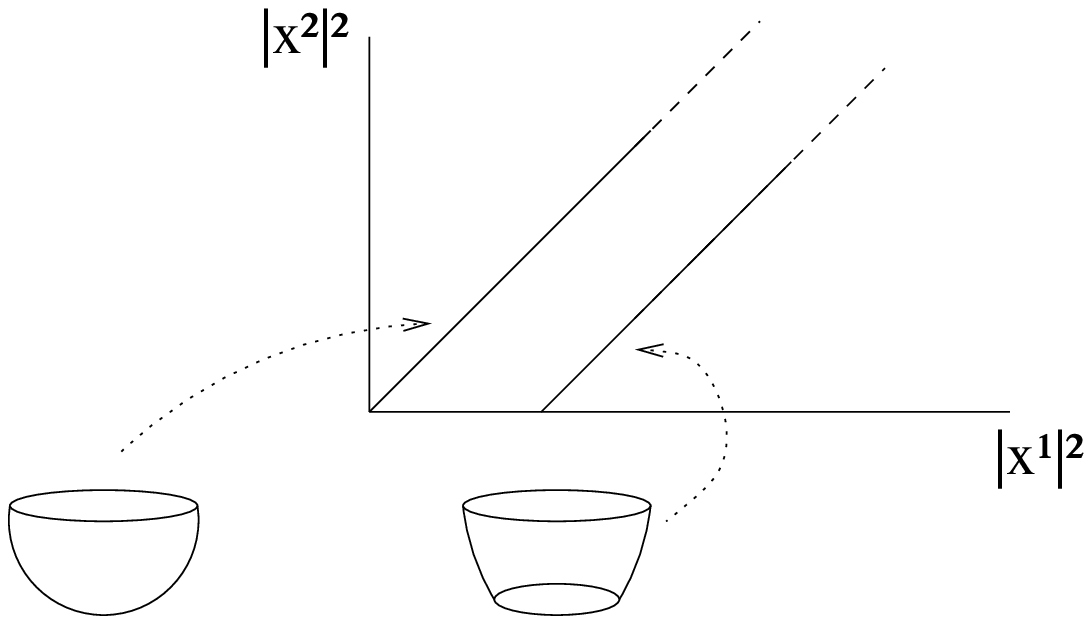}}
\rightskip 2pc
\noindent{\ninepoint\sl \baselineskip=8pt {\bf Fig.1}: {\rm 
Lagrangian submanifolds of ${\bf C}^2$ with and without boundaries, projected
to the two dimensional base $L=(|x^1|^2,|x^2|^2)$.}}
\bigskip


\subsec{Calabi-Yau Geometry and Special Lagrangian Submanifolds}

So far we have discussed a very simple Calabi-Yau geometry, namely the
non-compact ${\bf C}^n$.  However, toric geometry is also very
useful in describing rather non-trivial Calabi-Yau manifolds, both as
non-compact weighted projective spaces or complete
intersection in products of weighted projective spaces. We
first review some aspects of these constructions and their
relation to linear sigma model.

Start again with $X={\bf C}^n$ as a torus $T^{n}$ fibration over $L$.
The torus acts on $X$ by phase rotations $x^i
\rightarrow e^{i \theta^i}x^i$ and this action preserves the Kahler form.
Naive quotients by subgroups of $U(1)^n$ are neither smooth
nor Kahler (or complex for that matter) but there is a well known
prescription that circumvents both problems.

Pick a  $G=U(1)^{n-k}$ subgroup of the isometry group acting on $X$
by
\eqn\qt{x^i \rightarrow e^{i Q^{a}_{i}\epsilon_{a}} x^{i}}
for some choice of charges $Q^{a}$. If we define the quotient $Y=X//G$
to be obtained by setting for $a=1,\ldots, n-k$
\eqn\higgs{\sum_i Q_i^{a}|x^i|^2 = r^{a}}
on $X$ and dividing the resulting space by $G$ than the quotient
manifold $Y$ is a complex, Kahler manifold. This definition has a
natural realization through linear sigma models \witls\ where one
considers a two-dimensional $N=2$ gauge theory with gauge group
$G=U(1)^{n-k}$ and $n$ fields $\Phi^i$ which have charges $Q_i^a$ 
under the corresponding $U(1)$'s.  The above constraint \higgs\ is 
the minimum of the D-term potential $D^a=0$
and modding out the resulting space by $G$ is considering the
gauge inequivalent orbits of the vacuum. 

For sufficiently generic choices of
parameters $r^a$, $G$ acts freely on \higgs \ and $Y$ is
a smooth manifold. The Kahler form $\omega_Y$ on the quotient is
obtained from the Kahler form $\omega$ on $X$ by restricting to 
$D^a=0$ subspace and dividing by $G$.  

$Y$ can be also be viewed as a (generalization of) weighted
projective space $Y=X/G^{\bf C}$, where instead of 
setting $D$-terms to zero and dividing by $G$ we take
an ordinary quotient by the complexified gauge group $G^{\bf C}$  
$$x^i \sim \prod_a {(\lambda_a)}^{Q_i^a} x^i$$
for $\lambda_a$ in ${\bf C}^*$ , and with suitable subspaces of $X$ 
deleted.
The manifold $Y$ is in addition to being Kahler,
a non-compact Calabi-Yau space if, for each $a$,
\eqn\cycond{\sum_{i=1}^n Q_i^a=0.}
Note that this requires having some negative charges $Q_i^a$ and
the corresponding fields lead to the non-compact directions of the
Calabi-Yau.
Under the above condition the holomorphic $n$-form
$\Omega=dx^1\wedge \ldots \wedge dx^n$ is $G^{\bf C}$ 
invariant and descends to a
holomorphic $k$ form on $Y$ by contraction with $n-k$ generators
of the complexified gauge group, $\Omega_Y = i_{g^1}\ldots
i_{g^{n-k}}\Omega$.

We have to clarify what we mean by the manifold $Y$ being a Calabi-Yau
space: It has a trivial canonical line bundle.  This does
not mean that the metric induced from its embedding in $X$ 
agrees with the Ricci-flat Calabi-Yau metric. In fact it does not.
However as discussed in \witls\ the linear sigma model with Kahler
form induced from $X$ is a quantum theory on the worldsheet which
flows in the infrared to a conformal theory with an approximately 
Ricci-flat metric
(note that generally the metric
picked by the conformal theory is a refinement
of the Ricci-flat metric on the CY which only at large radii
 becomes the Ricci-flat metric).  The RG flow affects the
D-term  and leaves the
superpotential terms unchanged -- which is why for issues
of topological strings, mirror symmetry
works equally well for this non-Ricci-flat induced metric.

We now turn to construction of Lagrangian submanifolds of $Y$,
which can be defined since $Y$ is Kahler.
First, note that the geometric picture with $X$ realized as a $T^n$
fibration over $L$ descends to the quotient space.
The manifold $Y$ is a $T^n/G = T^k$ fibration over 
restriction of $L=R^n$ to subspace \higgs \ determined
by charges $Q_a$. The restriction, which we will denote by $L_Y$,
is clearly Lagrangian in the induced Kahler form $\omega_Y$. 
In fact all the Lagrangian submanifolds of $X$ we constructed in the
previous section descend to Lagrangian submanifolds of $Y$.
Because the Kahler form on $Y$ derives from the one on $X$ 
by restriction modulo $G$ Lagrangian submanifolds on $X$,
provided they make sense in the quotient, are automatically 
Lagrangian on $Y$ as well. 

The condition we need to impose is that the
$v^i_{\beta}$ should lead to gauge invariant constraints in \lagfiber ,
and this means that
$$Q^a\cdot v_{\beta}= \sum_{i=1}^{n} Q_i^a v^i_{\beta}=0.$$
The gauging constrains $v^{i}_{\beta}$ but it
does $not$ put a constraint on the Lagrangian charges $q_i^{\alpha}$.
The $q$'s and $Q$'s, up to taking linear
combinations, are thus the data specifying 
homology class of Lagrangian submanifold
of $Y$.  

Again note that the same comment
made above about the exact metrics on the Calabi-Yau manifold applies
equally well to the Lagrangian subspaces. Namely the
Lagrangian submanifolds we have constructed here will
not necessarily be Lagrangian with respect to the exact
metric picked by the conformal theory.  However one
expects that the Lagrangian submanifold gets deformed in
the IR, just as the metric gets deformed, so as to continue
to be Lagrangian.  Again, as far as the issues
of topological strings are concerned these are D-term
variations which do not affect the topological computations.
 

Given a Calabi-Yau manifold, one can formulate the condition for
Lagrangian submanifold to be of minimal volume in terms of the
holomorphic $n$-form $\Omega$. One defines a special Lagrangian
cycle to be that on which $\Omega$ has constant phase 
\hl \ref\syz{A.~Strominger, S.~Yau and E.~Zaslow,
``Mirror symmetry is T-duality,''Nucl.\ Phys.\  {\bf B479}, 
243 (1996) [hep-th/9606040].}.
%
%
If the Lagrangian submanifold satisfies this, it is volume minimizing
in its homology class.
In our case, $Y$ is Calabi-Yau if $\sum_i Q_i^a=0$. Since all the Lagrangian
submanifolds we constructed correspond to planar subspaces $D^r$ of $L_Y$ 
phase of $\Omega_Y$ on each is given by $\sum_i \theta^i$, so for our
constructions to lead to special Lagrangians this sum must be constant.
In order for the special Lagrangian condition
to be satisfied on $D^r$ without over-constraining the Lagrangian, 
we must have that of one the $v^{\beta}$ is $v^{\beta}_i =(1,1,1,...,1)$.
This in turn, by virtue of $q^{\alpha}\cdot v_\beta=0 $,
implies the constraint we stated before:
\eqn\consplaagain{\sum_i q_i^{\alpha}=0}
for all $\alpha$. From now on, we restrict our attention to Lagrangian
submanifolds which satisfy this.

We can also impose hypersurface constraints in $Y$ or consider complete
intersections in the weighted projective spaces. Physically this
corresponds to deforming the action of the two dimensional sigma model by
certain superpotential terms \witls . In these cases, the restriction of the
Kahler form of $Y$ to the corresponding subspaces gives a
Kahler structure to the Calabi-Yau. Thus the
intersection of the Lagrangian submanifolds we have constructed
with the Calabi-Yau manifold, continue to be Lagrangian.

To summarize we have constructed, for non-compact or compact
Calabi-Yau, characterized by charges $Q_i^a$ of the fields
$\Phi_i$ of the linear sigma model, Lagrangian submanifolds
characterized by ``charges'' $q_i^{\alpha}$.  These are
special Lagrangian if and only if \consplaagain\ holds.  These subspaces
are Lagrangian relative to the induced Kahler form from their
embedding in ${\bf C}^n$.  Moreover, they are expected to flow
in the IR
to special Lagrangian submanifolds
relative to the Kahler form corresponding
to the metric which gives rise to a conformal
theory on the worldsheet (and which at very large radii 
is close to the Calabi-Yau
metric).
For the sake of a shorter terminology
when we consider D-branes wrapped around such special
Lagrangian submanifolds we will refer to them as ``A-branes''
(as they preserve the A-model topological charge).

\newsec{Mirror Symmetry Action on Lagrangian D-branes}

In this section we obtain the mirror of the Lagrangian
D-branes constructed in section 2.  We first review the
derivation of mirror symmetry \hova \ and then use it
to find the ``B-branes'' that are mirror of the ``A-branes''.  We will
mainly concentrate on the Calabi-Yau case, and D-branes wrapped
over the special Lagrangian submanifolds-- however
many of our remarks apply to more general settings
including the non-Calabi-Yau cases.

Consider, for definiteness, 
a linear sigma model with fields $(\Phi_i,P)$ where
$i=1,...,n$ charged under a $U(1)$ with charges given by
$(Q_i,Q)$ \foot{In this section for convenience
we have shifted
our notation from the previous section in that we have $n+1$ total
fields rather than the $n$ fields of the previous section.}.
The Calabi-Yau condition
(equivalently the vanishing of the beta function) requires
$$Q+\sum_i Q_i =0$$
which implies that at least some of the charges are negative.  Let us
suppose that $Q<0$. The above equation is equivalent then to
$$|Q|=\sum_i Q_i$$
There is a potential in the linear sigma model which comes from
the D-term, and the minimum of this potential is given by
\eqn\dterm{\sum Q_i |\phi_i|^2 + Q |P|^2=r.}
The $r$ parameter is a FI term which combines with the $U(1)$
$\theta$ angle to give a complexified Kahler parameter $t=r+i
\theta$. When $r>0$ the geometry of this minimum modulo gauge
transformation can be viewed as a non-compact weighted projective
space with weights given by $(Q_i,Q)$. The Kahler class of the 
compact part of the space depends linearly on $r$, and the 
non-compact direction is parameterized by the field $P$.

To obtain the mirror model we follow  \hova\ and introduce dual (twisted)
chiral fields $Y_i$ such that
$${\rm Re} Y_i= |\Phi_i|^2$$
\eqn\veim{{\rm Re} Y_P= |P|^2.}
This is obtained by acting with
T-duality on all of the $n+1$ fields of the original theory.\foot{The
proposal for studying the geometry of the mirror Calabi-Yau in
terms of mirror symmetry action on tori \syz\ 
also uses T-duality, but in a different set up.  For example
for the case of quintic the approach of \hova\ applies 
T-duality to
6 fields, whereas in the \syz\ approach one applies it to 3 fields.
The approach of \syz\ is related to the 
 heuristic derivation of Batyrev's proposals for mirror pairs
given in \ref\leungv{N.~C.~Leung and C.~Vafa, ``Branes and toric
geometry,'' Adv.\ Theor.\ Math.\ Phys.\  {\bf 2}, 91 (1998)
[hep-th/9711013].}.  For some recent progress in this direction 
see for example \ref\ruan{W-D. ~Ruan, ``Lagrangian torus fibration
and mirror symmetry of Calabi-Yau hypersurface in
toric variety'', math.DG/0007028.}. 
However all approaches to understanding mirror symmetry have the
common feature of using T-duality in one way or another.}
It is also convenient to define
$$y_i={\rm exp}(-Y_i), \qquad y_P={\rm exp}(-Y_P) $$
and this is natural given the fact that the imaginary part of $Y_i$
are periodic variables, of period $2 \pi $. Moreover the mirror
version of the equation \dterm\ is given by
\eqn\bcons{y_p^Q\prod y_i^{Q_i}=e^{-t} \rightarrow \prod
y_i^{Q_i}=e^{-t}y_p^{|Q|}.}
The mirror theory is a Landau-Ginsburg theory in terms of $Y_i,P$ with a
superpotential
$$W=\sum_i y_i + y_p$$
subject to \bcons .  For simplicity, let us assume that all $Q_i$
divide $|Q|$ and put $m_i=|Q|/Q_i$. We then can solve
\bcons\ by introducing new fields ${\tilde y}_i^{m_i}=y_i $ in terms of which
we have
$$W=F({\tilde y}_i)=\sum_i {\tilde y}_i^{m_i}+ e^{t/|Q|}\prod_i {\tilde
y}_i.$$
To be precise, for the new fields to be well defined functions of the old, 
we have to consider an orbifold of this
acting on ${\tilde y}_i$ by all $m_i$'th roots of unity which
leave $\prod {\tilde y}_i$ invariant. 

Mirror symmetry above can also be stated in the geometric language.
We first recall the compact Calabi-Yau case.
The original theory reduces to a compact Calabi-Yau
sigma model if we add a gauge invariant superpotential
$PG(\phi_i)$. As discussed in \hova\ this does not affect
the LG superpotential $W$ above, 
except to make the fundamental fields of the theory to be $y_i$
instead of the $Y_i$. Then, the LG theory is equivalent to an
orbifold of the hypersurface
$$F({\tilde y_i})=0$$
in the corresponding weighted projective space. This can be
written in a coordinate patch where, say ${\tilde y}_n \neq 0$ as
$$F({\widehat y_i},{\widehat y_n}=1)=0$$
in inhomogeneous variables ${\widehat y}_i={\tilde y_i}/{\tilde y_n}$.

If in the original theory
we do not add a superpotential $PG(\Phi_i)$, then the $A$-model 
continues to correspond to a non-compact Calabi-Yau space. In this case
the mirror theory is geometrically equivalent \hiv\ 
to a non-compact Calabi-Yau
\eqn\nonc{xz=F({\tilde{y}}_i)}
where $x,z$ are ${\bf C}$-valued and ${\tilde{y}}_i\in {\bf C}^*$ (i.e.
in this case the $Y_i$ are the good variables). 
There is still a ${\bf C}^*$ action
on the above space, which allows us to set one of the ${\tilde{y}}_i$
to 1 (which one we set to one, depends
on the patch we wish to study the mirror geometry in). Note that the
non-compact case has two dimensions more compared to the compact
case (given by the extra variables $x,z$) but both the compact and the
non-compact geometry are characterized by $F$.

To avoid unnecessary complication in notation, in the following
we will drop the tilde off of ${\tilde y}_i\rightarrow y_i$.
Generalization of the above discussion to multiple $U(1)$'s is
straightforward and can be found in \hova .

\subsec{Identification of the B-branes} The mirror of Lagrangian
submanifolds are expected to be holomorphic submanifolds of the
mirror, which we call B-branes. Note that the action of T-duality
on ${\bf C}^n$ already suggests that the mirror of $ D^r$, whose
fiber is a $T^{n-r}\subset T^n$, is a $T^r$ fibration over $D^r$
i.e. it should be specified by $n-r$ complex equations.  This we
will indeed find to be the case.

In the discussion of special Lagrangian submanifolds we noted that
they are characterized by certain ``charges'' $q_i^{\alpha}$.
These in particular restrict the $\Phi_i$ by
$$\sum_i q_i^{\alpha }|\Phi_i|^2 =c^{\alpha}$$
with no loss of generality we have assumed that $q_P^{\alpha }=0$
(we can use the \dterm\ constraint to write the equations without $P$). 
Note also that the condition of being special Lagrangian submanifold 
implies that  $\sum_iq_i^{\alpha}=0$. 
Given the discussion above, it is easy to
write the mirror of the above Lagrangian.  Namely, from \veim\ and from
the fact that we expect a holomorphic equation we immediately find
that
\eqn\mirap{ \prod_{i=1}^{n-1} y_i^{q_i^{\alpha} \cdot m_i}
=\epsilon^{\alpha} {\rm exp}(-c^{\alpha}),}
where $\epsilon^{\alpha}$ is a phase, which can be 
combined with $c^{\alpha}$ to
give it an imaginary part. This implies that in the compact case
in addition to $F(y_i)=0$ we consider the holomorphic subspace
given by the above equations. For the non-compact case the same
holds, but for the subspace of $xz=F(y_i)$.

Below we give some examples of the mirror action on the A-branes
leading to B-branes on the mirror manifold for both compact and
non-compact cases.

\subsec{Compact Examples} Consider the quintic three-fold
as an example.  The field content
of the linear sigma model is a $U(1)$ gauge theory with six fields
with charges
$$(\Phi_1, \Phi_2, \Phi_3 ,\Phi_4 ,\Phi_5 ,P)=(1,1,1,1,1,-5)$$
(together with a superpotential defining
the complex structure of the quintic).  The mirror theory is given
by
\eqn\mirqu{[y_1^5+y_2^5+y_3^5+y_4^5+y_5^5+e^{t/5}y_1y_2y_3y_4y_5
=0]/\Gamma}
in $CP^4$, where we consider a $\Gamma=Z_5^3$ orbifold of it given by
multiplication of each $y_i$ by a fifth root of unity, preserving
$y_1y_2y_3y_4y_5$.

Now consider the mirror of the Lagrangian submanifold defined by
the charge $q_i^1$ given by
$$q^1=(1,-1,0,0,0,0).$$
This means that the Lagrangian submanifold
satisfies
$$|\Phi_1|^2-|\Phi_2|^2=c^1.$$
Then according to \mirap\ the mirror is given by the subspace of
\mirqu\ satisfying
\eqn\firs{y_1^5=y_2^5 {\rm exp}(-c^1).}
This is a two complex dimensional holomorphic subspace.

As another example, consider the Lagrangian submanifold given by
two charges $q^1,q^2$ with $q^1$ as given above, and
$$q^2=(0,0,1,0,0,-1)$$
which means that we are imposing that the Lagrangian submanifold 
intersects the base at
$$|\Phi_3|^2-|P|^2=c^2.$$
As mentioned before, we can change this (by imposing the condition
of the D-terms)
$$|\Phi_1|^2+|\Phi_2|^2+|\Phi_3|^2+|\Phi_4|^2+|\Phi_5|^2-5|P|^2=r$$
to
$$-|\Phi_1|^2-|\Phi_2|^2+4|\Phi_3|^2-|\Phi_4|^2-|\Phi_5|^2=
5c^2-r={\tilde c}^2$$
where we have introduced ${\tilde c}^2$ for convenience. In other
words this is effectively equivalent to taking
$q^2=(-1,-1,4,-1,-1,0)$.  This leads to the mirror brane given as
the locus characterized, in addition to  the constraint \firs\ by
$$(y_3^5)^4=y_1^5y_2^5 y_4^5 y_5 ^5 e^{-c_2}$$
which is a complex dimension one subvariety of the mirror to
quintic.

\subsec{Non-compact Examples}

As our first non-compact example we consider the geometry given by
the $O(-1)+O(-1)$ bundle over ${\bf P}^1$. This is described by a
$U(1)$ linear sigma model with four fields with charges
$$(\Phi_1, \Phi_2 , \Phi_3 , \Phi_4 )=(1,1,-1,-1).$$
The mirror of this theory is given by the geometry
\eqn\loccsu{xz=y_2+y_3+y_4+e^{-t}{y_3y_4\over y_2}}
where $x,z \in {\bf C}$ and $y_2,y_3,y_4$ are ${\bf C}^*$
variables, and we have to go to a patch where one of the $y_i=1$
(we have eliminated $y_1$ from the superpotential
by the equation $y_1y_2=y_3y_4e^{-t}$, as we will be
in a regime of parameters where $y_1$ is small and varies little).
The convenient choice of patch for the A-branes we will consider turns
out to be given by $y_4=1$, in which case the equation of the mirror is
\eqn\locsu{xz=y_2+y_3+1+e^{-t}{y_3\over y_2}.}
We consider the A-brane characterized by two charges
$$q^1=(0,1,0,-1)$$
$$q^2=(0,0,1,-1)$$
which corresponds to the projection on the base given by
$$|\Phi_2|^2-|\Phi_4|^2=c_1$$
$$|\Phi_3|^2-|\Phi_4|^2=c_2$$
and consequently $|\Phi_1|^2-|\Phi_4|^2=r -c_1+c_2$.
This makes sense for generic $c_1,c_2$
see figure 2\foot{For a geometric meaning of such figures
as well as an interpretation in terms of branes
see  \leungv\ .}. 
However, as noted before, there are certain codimension one
loci in parameter space where something special happens:
The generic Lagrangian submanifold, corresponding to case III in figure 2, 
degenerates to two submanifolds and we can in principle 
wrap the D-brane over
any one of them. This happens for example if (I) $r>c_1>0$ and $c_2=0$ or if
(II) $c_1=0$ and $c_2>0$ (see figure 2). 
For either of these two cases the D-brane will {\it not} have a
deformation away from this special locus, as it would acquire a boundary.  
Precisely for these branes we will
later compute a non-vanishing superpotential using the mirror
B-brane.
\bigskip
\centerline{\epsfxsize 4.0 truein\epsfbox{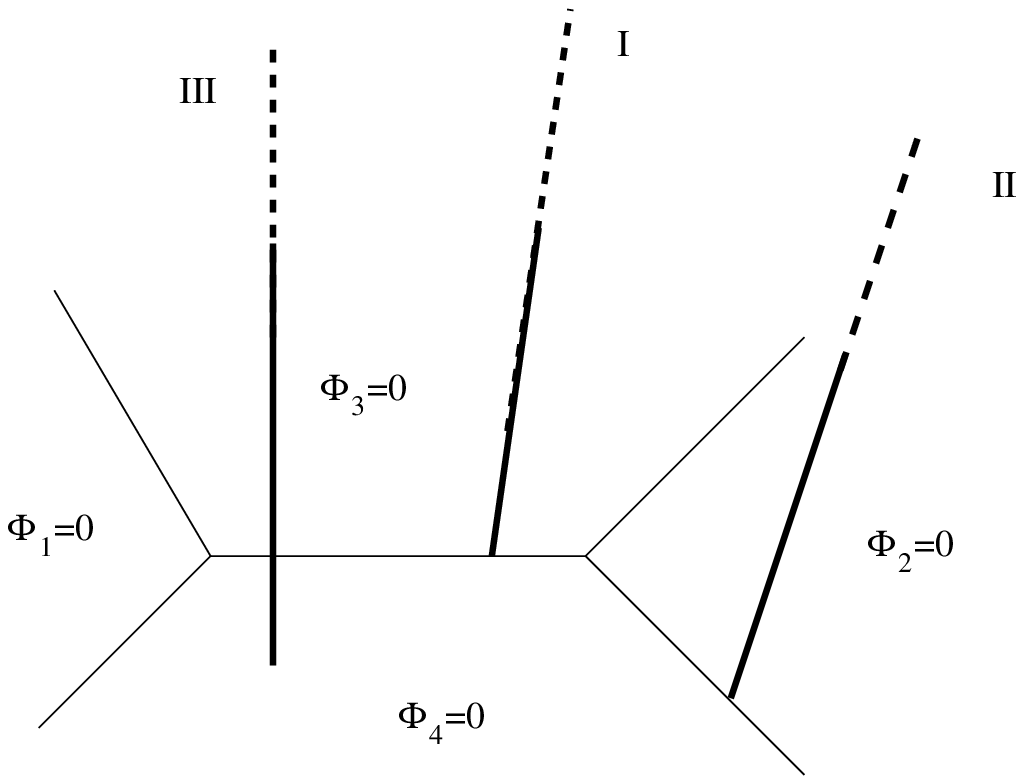}}
\leftskip 2pc
\rightskip 2pc
\noindent{\ninepoint\sl \baselineskip=8pt {\bf Fig.2}: 
{\rm The projection of the Lagrangian
submanifold on the base corresponds to a straight
line (III).  For special values of $c_1,c_2$ the
line will intersect the loci with a pair of vanishing circles.
This can happen in two inequivalent ways. For $r>c_1>0,~c_2 =0$ it ends
on the interval $\Phi_3=\Phi_4=0$ (I) and if
$c_1=0$ it ends on the line $\Phi_4=\Phi_2=0$ (II).}}
\bigskip

The mirror for the general values of $c_1,c_2$ is given by the
subspace of \locsu\
$$y_2=e^{-c_1}y_4 \qquad y_3=e^{-c_2} y_4.$$
This implies that in the $y_4=1$ patch using \locsu , we look at the subspace
\eqn\relev{y_2=e^{-c_1} \qquad y_3=e^{-c_2} \qquad {\rm of}\qquad
xz=1+y_2+y_3+e^{-t}{y_3\over y_2}.}
Note that this subspace is given by a
one-dimensional complex B-brane characterized by
$$xz={\rm const.}$$
Note that if the constant on the RHS is zero, then the B-brane
splits to two B-branes given by $x=0$ or $z=0$.  This is
the mirror of the statement we made about the A-brane. Let us
check this for the two cases mentioned above in the large radius
limit, where the two pictures should match.

Consider first the case I where $c_2=0$ and where we consider
the large radius limit $r>>0$ and where $c_1$ is large but
less than $r/2$
(i.e. when the A-brane intersects the ${\bf P}^1$ near the
equator and towards the south pole).
 In this limit the RHS is dominated by $1+y_3$
and if we take the imaginary part of
$c_2$ (which was not fixed by the mirror map) to be $i \pi$ we see
that for this brane $y_3=-1$ and the RHS vanishes.
Thus the mirror of the half A-brane agrees in this limit with the
locus where $xz=0$ as expected.  The generalization of this
condition is predicted by the mirror map to be choosing
$y_3$ as a function of $y_2$ such that the RHS vanishes  
away from the large radius limit. Writing
in terms of the ${\bf C}^*$ variables $y_2=e^u$ and $y_3=e^v$, this means that
we can determine $v$ as a function of $u$ such
that $F(u,v)=0$ where
$$xz= F(u,v)=1+e^{u}+e^{v}+e^{-t}e^{(v-u)}.$$
To leading order $v={i \pi}$, but more generally we have that
\eqn\thre{v=i\pi +log{{1+ e^u}\over{1 + e^{-t-u}}},}
as implied by $F(u,v)=0$.
Note that here $u$ geometrically denotes the size of
the disc in the ${\bf P}^1$ which ends on the brane.
This is the sense, as we will discuss later, in which 
$u$ is the ``good variable'' from the viewpoint
of topological string.

In the case (II) where we consider $c_1=0$ and $c_2>0$,
in the large radius limit we have  $e^{u}=-1$, $e^v \rightarrow 0$ 
and again the RHS of the equation $xy=F(u,v)$
vanishes. More generally, i.e. away
from the large radius limit, to obtain the mirror of the
single brane we demand vanishing of $F$ which in this
case gives
\eqn\anot{u=i\pi+log[{{1+e^v}\over{2}}+ {{1}\over{2}} \sqrt{{(1+e^v)}^2 -
4e^{-t+v}}] .}
Note that in this case $v$ is the good variable,
as it measures the size of the
disc passing through the south pole of ${\bf P^1}$.

For another example, consider the local model given by a
non-compact Calabi-Yau containing a ${\bf P}^1\times {\bf P}^1$.  This can
be realized with a $U(1)^2$ gauge group with five matter fields,
with charges
$$Q^1=(1,1,0,0,-2)$$
$$Q^2=(0,0,1,1,-2)$$
the mirror manifold in the $y_5=1$ patch is given by
\eqn\rer{xz = y_1+{e^{-t_1}\over y_1} + y_3 + {{e^{-t_2}}\over {y_3}} +1.}
 We consider the Lagrangian
submanifold given by
$$q^1=(1,0,0,0,-1)$$
$$q^2=(0,0,1,0,-1),$$
which means that we have put 
$$|\Phi_1|^2 = |\Phi_5|^2 +c_1$$
$$|\Phi_3|^3=|\Phi_5|^2+c_2.$$
%

\bigskip
\centerline{\epsfxsize 4.0 truein\epsfbox{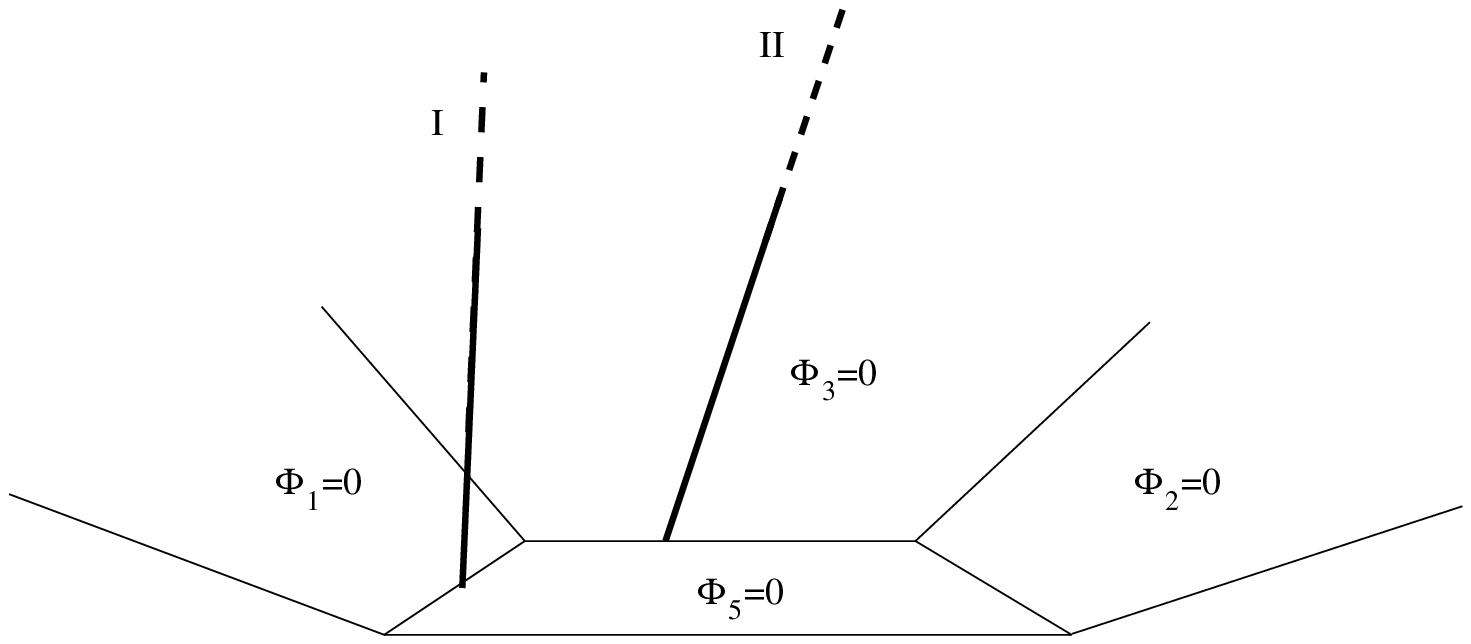}}
\leftskip 2pc
\noindent{\ninepoint\sl \baselineskip=8pt {\bf Fig.3}: 
{\rm Some special A-branes in $\IP^1 \times \IP^1$.}}
\bigskip

The mirror is given by
$y_1=e^{-c_1}y_5$ and $y_3=e^{-c_2}y_5$, or going to
the $y_5=1$ patch, by $y_1=e^{-c_1}$ and $y_3=e^{-c_2}$
subspace of \rer\ .  If we substitute $y_1=e^{u}$, $y_3 =e^v$
into \rer , we get an equation of form  $xz=F(u,v)$.
The condition that the brane splits to two parts is again the
condition that is quantum corrected to $F(u,v)=0$.
For example in the large radius limit if we consider
$0<<c_1 < r/2$ and $c_2=0$ we have the brane II
depicted in figure $3$.
The mirror brane is $e^u = e^{-c_1}\rightarrow 0$, $e^v=-1$ (by
a suitable choice of imaginary part of $c_2$) and so $F=0$
is satisfied. More generally, we have $v$ determined 
in terms of $u$ from
$F(u,v)=0$.

\newsec{Topological Strings and Superpotentials}

In the previous sections we have considered certain special Lagrangian
submanifolds in Calabi-Yau manifolds and their mirrors,
the holomorphic submanifolds of the mirror geometry.  This statement
descends to the topological subsector of these theories.
In particular, topological A-models admit Lagrangian D-branes (which
is why we called them A-branes) and topological B-models admit
holomorphic D-branes (and hence the terminology B-branes) 
\ref\witcss{E.~Witten,``Chern-Simons gauge theory as a string theory,''
hep-th/9207094.}.  Since mirror symmetry
converts the A-type topological string to B-type topological string,
and the A-branes to B-branes, it is natural to ask how one can
use mirror symmetry to compute A-type topological string invariants
in terms of the B-model.  This general setup
and its consequences for topological strings has been discussed in
\ref\cext{C.~Vafa,``Extending mirror conjecture to Calabi-Yau with bundles,''
hep-th/9804131.}.

The A-model topological string amplitudes are given
in terms of the enumerative geometry having to do
with holomorphic maps from Riemann surfaces with boundaries
to a target Calabi-Yau manifold where the boundary ends on 
a Lagrangian D-brane\foot{The degenerate
limit of such maps gives rise to ordinary
Chern-Simons theory on the Lagrangian submanifold \witcss .}.
  This in general involves a complicated
enumerative geometry question and there is no direct approach
known to computing it.  
Using the large $N$ duality conjecture \ref\gopva{R.~Gopakumar 
and C.~Vafa,``On the gauge theory/geometry correspondence,''hep-th/9811131.}\
there have been some cases where one can compute certain corrections
involving holomorphic maps from Riemann surfaces
with boundaries to target space geometry
\ov \ref\mala{J.~M.~Labastida and M.~Marino,
``Polynomial invariants for torus knots and topological strings,''
hep-th/0004196.}\ref\malava{J.~M.~Labastida, M.~Marino and C.~Vafa,
``Knots, links and branes at large N,''JHEP {\bf 0011}, 007 (2000)
[hep-th/0010102].}.
Moreover based on what the topological strings compute in the
context of type II superstrings certain integrality
properties for the A-model amplitude can be predicted \ov\
generalizing those without D-branes 
\ref\govabu{R.~Gopakumar and C.~Vafa,``M-theory and topological 
strings I,II'', hep-th/9809187, hep-th/9812127.}.
For example it is shown that the disc amplitudes in the A-model
will have the general structure given by
\eqn\pf{F_{Disc}=W =\sum_{n=0}^{\infty}\;\; \sum_{\vec{m},\vec{k}}
\frac{d_{\vec{k},\vec{m}}}{n^2} q^{n\vec{k}}y^{n\vec{m}},}
where $q = e^{-t}$ for $t$ a basis for complexified closed string Kahler
classes, and $y$ related by exponentiation to the complexified
open string Kahler class which measures the volumes of holomorphic
discs. The integers $d_{\vec{k},\vec{m}}$ in the above formula
count ``primitive disc instantons'' in relative homology class
$(\vec{m},\vec{k})$, where ${\vec{m}}$ labels the class on the boundary
(i.e. an element of of $H_1$ of the brane) and ${\vec{k}}$ labels an
$H_2$ element of the Calabi-Yau,
and the sum over $n$ sums the multi-coverings of
these.  The reason we have denoted the disc amplitude 
also by $W$ is that in the context
of type II superstrings, if we consider the branes filling the
space-time (which only makes sense if the Calabi-Yau is
non-compact, for the brane flux to have somewhere to go) the
topological string disc amplitude has the interpretation of
superpotential
corrections to 4d $N=1$ supersymmetric theory
\ref\bcov{M.~Bershadsky, S.~Cecotti, H.~Ooguri and C.~Vafa,
``Kodaira-Spencer theory of gravity and exact results for quantum 
string amplitudes,''Commun.\ Math.\ Phys.\  {\bf 165}, 311 (1994)
[hep-th/9309140].}\ref\do{I.~Brunner, M.~R.~Douglas, A.~Lawrence and 
C.~Romelsberger,``D-branes on the quintic,''JHEP {\bf 0008}, 015 (2000)
[hep-th/9906200].}\ov 
\ref\kket{S.~Kachru, S.~Katz, A.~Lawrence and 
J.~McGreevy,``Open string instantons and superpotentials,''
Phys.\ Rev.\  {\bf D62}, 026001 
(2000)[hep-th/9912151].}\ref\katzet{S.~Kachru, S.~Katz, A.~Lawrence 
and J.~McGreevy,
``Mirror symmetry for open strings,''Phys.\ Rev.\  {\bf D62}, 
126005 (2000)[hep-th/0006047].}.
Note that the above form of $W$ makes sense only in the
large radius limit and that this structure requires $W$ to have very
strong integrality properties.

On the B-model side the topological
string is related to holomorphic Chern-Simons theory \witcss\ 
if we consider the D-brane wrapped over the entire Calabi-Yau,
or its dimensional reductions depending on the dimension
of the D-brane (as we will discuss below).  Thus the hope is to map
the difficult problem of computations on the A-model side to some
easy computations on the B-model side.  For example, if we consider
an annulus, then the B-model partition function is given
by a holomorphic Ray-Singer torsion and this would compute, by
mirror symmetry holomorphic maps from the annulus to the original
Calabi-Yau geometry with the boundaries of the annulus ending
on the mirror Lagrangian submanifolds \cext .   Similarly
higher genus Riemann surfaces with boundaries have interpretation
in terms of the holomorphic Chern-Simons theory coupled
to the bulk complex structure (the Kodaira-Spencer theory 
\ref\bcov{M.~Bershadsky, S.~Cecotti, H.~Ooguri and C.~Vafa,
``Kodaira-Spencer theory of gravity and exact results for quantum string 
amplitudes,''Commun.\ Math.\ Phys.\  {\bf 165}, 311 (1994) [hep-th/9309140].}).

The disc amplitude computes the classical action
on the B-model side, which as noted above corresponds
to the holomorphic Chern-Simons action or its reductions
on the worldvolume of the B-brane.  Thus by computing
the classical action on the B-model side, we can compute
the A-model holomorphic disc instantons.  We will
use this idea to compute, using mirror symmetry, the
A-model disc instanton corrections.

\subsec{B-model Computation of Superpotential for a 2-brane}

Consider a Calabi-Yau manifold $Y$ in the context of topological
B-model. If we have a 6-brane wrapping the entire $Y$, which can
be viewed as introducing an open string sector with purely
Neumann boundary conditions on $Y$, we obtain a holomorphic
Chern-Simons gauge system
living on the brane, which in this case happens to be $Y$ itself,
with action given by \witcss\
\eqn\holcs{W =\int_{Y} \Omega \wedge 
{\rm Tr}[ A \wedge \bar{\partial} A +{2\over 3}
A\wedge A\wedge A].}
If we have $N$ branes, $A$ is a holomorphic $U(N)$ gauge connection
which can be viewed as a $U(N)$ adjoint valued $(0,1)$ form on $Y$.
For lower dimensional B-branes one obtains the reductions
of this action to lower dimensions. 
For example, for 0-branes all directions of $A$ become scalar\foot{
In this case the reduction agrees with the result in \do \
for the 0-branes superpotential
where the above action becomes $\Omega_{ijk}tr [\Phi^i,
\Phi^j]\Phi^k$.}.  

Here we are interested in the case where the
B-branes are two real dimensional (i.e. one complex dimensional) so wrap
curves $\cal{C}$ in $Y$. We
restrict attention to the case of a single 2-brane
and consider the reduction of the holomorphic Chern-Simons theory
to $\cal{C}$. 

Restricted to
$\cal{C}$ the tangent space $T_Y$ of the Calabi-Yau $Y$ splits as
$$T_p(Y)=T_p({\cal C })\oplus N_p({\cal C})$$
where $T_p({\cal C})$ denotes the tangent directions to ${\cal C}$ and 
$N_p({\cal C})$ denotes the normal directions at a point $p$ on  ${\cal C}$.
Two directions of the gauge field $A$ give two independent
sections of the normal bundle 
$N({\cal C})$, we denote them $\phi^{i}$, $i=1,2$. They should be viewed as 
deformations of $\cal{C}$ in $Y$.  

Since the canonical bundle of $Y$ is
trivial, it implies that $\wedge^2 N(\cal C)$ can be identified 
with $T_{\cal C}^*$,
and the identification is done via contraction with
the holomorphic 3-form $\Omega$ restricted
to ${\cal C}$.  In other words, we have the pairing
$$U_z=\Omega_{ijz} \phi^i \wedge \phi^j$$
where $z$ denotes a coordinate system on ${\cal C}$.  Using this, it
is straightforward to write the dimensional reduction of holomorphic
Chern-Simons theory on ${\cal C}$ which is given by
\eqn\hocc{W({\cal C})=\int_{\cal C} \Omega_{ijz} \phi^i {\overline \partial_z}
 \phi^j dzd{\overline z}.}
Here we are using a coordinate system on $Y$ 
where $\Omega_{ijz}$ is a constant,
as can always be done on a Calabi-Yau three-fold.

\subsec{Another Reformulation of the Superpotential Computation}

Note that locally we can write the closed 3-form $\Omega$ as 
$$\Omega =d \omega$$
in particular $\Omega_{ijz}=\partial_z \omega_{ij}\pm {\rm perm}$.
Using this and integrating by parts we can rewrite
\hocc\ as
$$W({\cal C})=\int_{\cal C} \omega$$
where here by ${\cal C}$ we mean any of the curves 
arising by deformations of the base curve by the sections of the normal
bundle $\phi^i$. 
Note that, even though $\omega$ is not globally well defined in general,
the above action $W(\cal{C})$ is well defined, at least 
as long as ${\cal C}$ has no
boundary.

We can now reformulate the superpotential computation
in a way which makes contact with another, space-time, viewpoint
\ref\wittsupe{E.~Witten,``Branes and the dynamics of QCD,''Nucl.\ Phys.\ 
Proc.\ Suppl.\  {\bf 68}, 216 (1998).}\ref\dtho{Donaldson, S.K.
 and Thomas, R.P. (1998).
{\it {Gauge theory in higher dimensions}} In: The Geometric Universe:
Science, Geometry and the work of Roger Penrose, S.A. Huggett et al
(eds), Oxford University Press.}, and which we will present in a 
slightly different form below.
This approach has been dicussed in the present context in \kket .

Consider type IIB superstring on a non-compact Calabi-Yau with 
a domain wall made of a $D5$ brane.  In $x<0$, the 5-brane
wraps over the cycle ${\cal C}$ and fills the spacetime. At $x=0$
it is the three chain $D$ times the $2+1$ dimensions
of spacetime and at $x>0$ it wraps over ${\cal C}_*$ and 
fills the spacetime again.  Then the BPS tension for this domain wall
is given by the ``holomorphic volume'' of $D$ which is
$\int_D \Omega$, and this should correspond to the change
in the value of the superpotential from left to right, which is given by
$W({\cal C})-W({\cal C}_*)$. Indeed,
\eqn\funfo{W({\cal C})-W({{\cal C}_*})=\int_{{\cal C}}\omega -\int_{{\cal C}_*}
\omega=\int_D\Omega}
where $D$ is a 3-chain with $\partial D ={\cal C}-{\cal C}_*$.  

Note that if we consider a family of ${\cal C}$ which
is holomorphic, then $W=0$.  One way
to see that is to use \hocc\ where it is clear that
if $\phi^i$'s are holomorphic functions of $z$, i.e. they correspond to
a holomorphic deformation of ${\cal C}$, then the superpotential vanishes.
Another way to see this is to use \funfo\ and note that
$\Omega$, which is a $(3,0)$ 
form restricted to a holomorphic curve ${\cal C}$ vanishes.
In \kket\ some non-vanishing superpotentials
were obtained by considering a family of curves with
obstructed holomorphic deformations, thus giving a non-vanishing $W$.   
In our application we find another way $W$ can be non-zero, and
that involves considering non-compact ${\cal C}$.  Fixing the boundary
condition at infinity can provide an obstruction for having
a holomorphic deformation of ${\cal C}$ and lead to a 
non-vanishing superpotential.

In order to do this we will need to apply \hocc\ to
manifolds ${\cal{C}}$ which are non-compact and in these
cases, in order to fix
the superpotential, we would need to know the boundary conditions on the
fields at infinity (which
will fix the total derivative ambiguities
of the action).  
This will be discussed later in the context
of examples.  
%
%

\subsec{B-brane superpotentials}

In this subsection we compute the superpotential for some of the $B$-branes in
non-compact Calabi-Yau three-folds $Y$ considered in section 2 as the mirror
of certain $A$-branes in the mirror non-compact Calabi-Yau.

Consider Calabi-Yau manifold $Y$ given by
$$xz=F(u,v)$$
where $F(u,v)=0$ is the equation of complex curve $\Sigma$,  given by
a polynomial in single valued variables $e^{u},e^{v}$ (recall
that $u,v$ are cylinder-valued).  The appearance of a Riemann surface
$\Sigma$ is familiar from the viewpoint of the $N=2$ Seiberg-Witten
geometry and their realization in terms of
 geometric engineering of $N=2$ theories using
type II strings propagating on a non-compact Calabi-Yau 3-fold.

We will compute the superpotential
for a D-brane wrapping the holomorphic curve $\cal{C}$ which is one
component of the collapsed fiber $xz = 0$. Concretely, we take
$${\cal{C}}~:~~~ x=0=F(u,v)~~~u=u_*~~v=v_*,$$
which leaves $z$ arbitrary and we identify it with a coordinate on
$\cal{C}$. Thus, $\cal{C}$ is
non-compact, of complex dimension one, and 
is parameterized by a point
of the Riemann surface $\Sigma$, denoted here by $u_*,v_*$.
We will now compute the  superpotential as a function of the
choice of a point $(u,v)$ on $\Sigma$ and relate it to Abel-Jacobi
map for a 1-form on $\Sigma$.

In order to do this, all we have to do is to compute the
brane action \hocc\ for the configuration
of the brane we are considering.  Our brane is parameterized
by $z$ and the two scalar fields of the theory on the brane can
be denoted by $u(z,{\overline z}),v(z,{\overline z})$, 
which represent its normal deformation inside the Calabi-Yau.
We fix the brane so that at infinity it approaches a fixed
$u_*,v_*$, i.e., 
$$u(z,{\overline z})\rightarrow u_*, \quad
v(z,{\overline z}) \rightarrow v_* \qquad {\rm as}
\qquad |z|>>\Lambda$$
for some fixed and large $\Lambda$.
To begin with we start with the brane for which $u,v$ are
identically equal to $u_*,v_*$ for all $z$.  We now want
to move the brane to a different value of $u,v$ on the Riemann
surface $F(u,v)=0$.  Since the boundary condition at infinity
is fixed this means that we can at most guarantee that
$u(z,{\overline z}),v(z,{\overline z})$ for  $|z|<\Lambda$
is fixed and equal to $(u,v)$, but that as $|z|\rightarrow \infty$
the $(u,v)$ go back to $(u_*,v_*)$.
In fact it is simplest if 
we consider a rotationally symmetric configuration of $u,v$
on the $z$ plane, so that $u,v$ do not depend on $\theta$,
and only depend on $r=|z|$ (see figure 4).
\bigskip
\centerline{\epsfxsize 4.0 truein\epsfbox{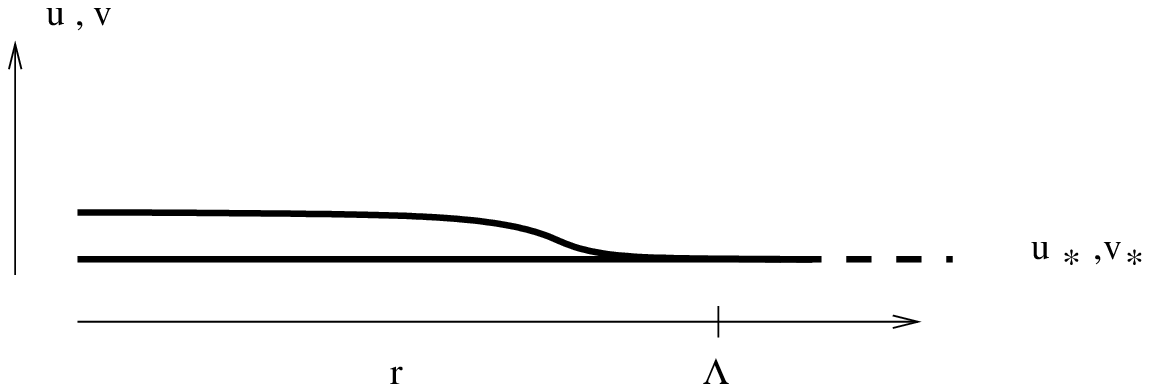}}
\leftskip 3pc
\rightskip 3pc
\noindent{\ninepoint\sl \baselineskip=8pt {\bf Fig.4}: {\rm 
Initial holomorphic brane configuration $(u(r),v(r))=(u_*,v_*)$
is deformed so that near the origin it is given by
$(u,v)$.}}
\bigskip

In writing the action \hocc , as noted before, in the non-compact
case we have to decide which field we keep fixed at infinity.
From the A-model side, as discussed before, there is always
a natural field corresponding to the
variable which measures the size of the disc instantons.
Let us say it is $v$ and we thus keep it fixed at infinity (which will of course fix
the {\it 
on-shell} value of $u$, by the condition that $F(u,v)=0$).
The holomorphic
three-form on $Y$ is $\Omega = du dv\frac{dz}{z}$, and
so using this we can write the action \hocc\ as
$$W=\int_{\cal C} {dzd{\overline z}\over z} u {\overline \partial_z}v $$
(Note that if $v$ were the field fixed at infinity
we would have written the action as $-v {\overline \partial_z}u$).
For the configuration at hand the $dz/z \sim d\theta $ integral
can be readily done and so we are left with the radial integral
in the $z$ plane which is given by
\eqn\lasteq{W({\cal{C}}) = \int_{v_*}^{v} u dv ~\rightarrow ~\partial_vW=u.}
The integral can be viewed as the integral of a 
1-form on the Riemann surface $F(u,v)=0$,
and we can view $u(v)$ determined by the condition
of being on the Riemann surface.  As mentioned
before, the superpotential can be viewed as an Abel-Jacobi map associated
to the 1-form $udv$ on the Riemann surface $F(u,v)=0$ where
the position of the brane is labeled by a point on the
Riemann surface.

So far we have talked about B-brane configurations
which are given by $x=0$.  We could have done the
same for the branes given by $z=0$.  The only difference
between them is given by a change in the sign of the superpotential
(because $dz/z =-dx/x$ and so the holomorphic 3-form
changes by an overall sign). Note that if we have a copy
of {\it both} kinds of branes, we can deform
the B-branes so that we are no longer on the Riemann surface
$F(u,v)=0$. Since the superpotential is the addition of these
two contributions it vanishes.

\newsec{A-brane Superpotential and Holomorphic Discs}

In this section we use the superpotential computation on the
B-model mirror  to compute holomorphic disc
instanton corrections to superpotentials of A-model branes for some 
of the examples discussed in this paper.  The idea is that
the disc amplitudes on the A-model side get mapped, by mirror
symmetry, to disc amplitudes on the B-model side, which
as we discussed in previous section, can be computed explicitly.
We will restrict our attention mainly on the A-model Lagrangian
submanifolds for which the B-model mirror predicts a non-vanishing
disc amplitude.  These correspond to the particular
class of Lagrangian
submanifolds that we discussed in section 3, 
which end on the `skeleton' of the toric
diagrams.

The right regime for the discussion is the limit where the A-model side is 
geometric, and that is the large radius limit for the Calabi-Yau.  
As far as the Lagrangian A-branes are concerned we should also
consider the regime of parameters where the discs that bound
the branes are large.  In this regime of parameters
we discussed some non-compact D branes in section 3
and they will serve as our main examples.  

As mentioned before, there are strong integrality
predictions for the disc amplitudes \ov :
\eqn\pf{W =\sum_{n=0}^{\infty}\;\; \sum_{\vec{m},\vec{k}}
\frac{d_{\vec{k},\vec{m}}}{n^2} q^{n\vec{k}}y^{n\vec{m}}}
where $q = e^{-t}$ for the complexified closed string Kahler
class $t$, $y$ is related by exponentiation to the complexified
open string Kahler class that measures the volumes of holomorphic
discs and $d_{\vec{k},\vec{m}}$ are integers. Below we
present some examples.  In the first
example we give, we recover the corresponding
answer predicted in \ov\ based on a completely different reasoning.
In the other examples we obtain more complicated results which
as we will discuss below pass the integrality check
in a non-trivial way.

\subsec{$O(-1)\times O(-1)$ bundle over ${\bf P}^1$}

We consider the small resolution of the
conifold given by $Q=(1,1,-1,-1)$ and the two charges
$q^1,q^2$ denoting the Lagrangian
submanifold discussed in section 3.  
There are two inequivalent ``phases'' for the
 Lagrangian submanifolds
that we will consider.  The two phases are denoted
by I and II in figure 2 and
we have already discussed, in section 3, how mirror symmetry
acts on them.  In particular,
in terms of the mirror variables
 $y_2/y_4 = e^{u}, y_3/y_4 =e^v$
the position of the brane is characterized
by $u,v$ subject to
$$0 = e^{-t}e^{v-u}+e^{u}+e^{v} +1.$$
As noted in section 3, in case I the natural variable from the A-model
perspective is $u$, and in the case II the natural
variable is $v$.

\noindent{\ninepoint\sl \baselineskip=8pt {{\cal{Phase}} I}}

In this phase $u$ is the physical field of the open string model, which
means that it measures the size of a minimal holomorphic disc ending
on the Lagrangian submanifold.

Since $u$ is the good variable, the superpotential is $W =- \int
v(u) du$, with $v(u)$ determined from the equation of the curve \thre .
For future convenience we use the freedom to shift 
the imaginary parts of the fields by $\pi$, and define new variables
$\hat{u}=u+i\pi,\hat{v}=v + i\pi$ (as discussed in section 3,
the value of the imaginary part is not fixed by mirror symmetry). 
In terms of shifted variables we have
$$\partial_{\hat{u}} W = -log{{1- e^{\hat{u}}}\over{1 - e^{-t-{\hat{u}}}}}.$$
This is in fact the expected answer \ov\ based on the
target space interpretation of topological string amplitudes.  To
see this it is convenient to factor out  $e^{-t-{\hat u}}$ from
the denominator which gives
$$W=P(t,{\hat u})+ \sum_{n>0}{{e^{n{\hat u}} - e^{n(t+{\hat u})}}\over
{n^2}}$$
where $P$ is a finite ambiguous polynomial in $t$ and ${\hat u}$.
This agrees with the result of \ov\ obtained by completely
different means where the two sums were also
interpreted in terms of the (multi-coverings) of two 
primitive discs 
wrapping the southern and northern hemispheres of the ${\bf P}^1$ and
ending on the Lagrangian
submanifold.\foot{To compare with \ov\ note that $-{\hat u}$ and
$t+{\hat u}$ are the two complexified areas of the two discs.}

\noindent{\ninepoint\sl \baselineskip=8pt {{\cal{Phase}} II}}

The natural variable for this phase is $v$ which measures
the size of the disc passing through the south pole and
ending on the Lagrangian submanifold (see figure 5).
\bigskip
\centerline{\epsfxsize 4.truein\epsfbox{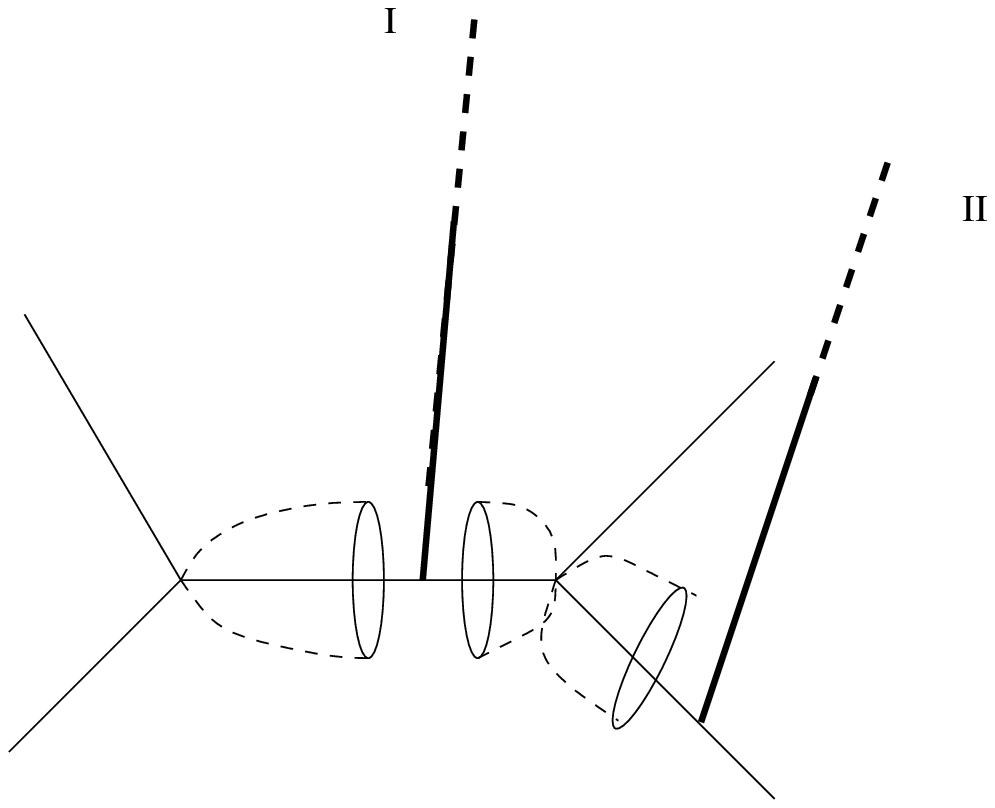}}
\rightskip 2pc
\noindent{\ninepoint\sl \baselineskip=8pt {\bf Fig.5}: {\rm Two
phases of the D-brane and some disc instantons.}}
\bigskip
 The superpotential is given by
\eqn\pot{\partial_{\hat{v}} W= \hat{u}(\hat{v}) =  
log(\frac{1-e^{\hat{v}}}{2}+
\frac{\sqrt{(1-e^{\hat{v}})^2+4e^{-t+\hat{v}}}}{2}).}
We can expand $W$ around the large radius limit as
$$\partial_{\hat{v}} W=
log(1-e^{\hat{v}})+ \sum_{k=1}^{\infty}\sum_{m=k}^{\infty} C_{k,m}e^{-k
t}e^{m{\hat  v}}$$
where, 
\eqn\coeff{ C_{k,m} =-\frac{(-1)^{k}}{k+m}B(k,k,m-k)}
and where
$$B(a_i)={(\sum_i a_i)!\over \prod_i a_i!}.$$
To check the integrality properties
of this amplitude \pf\ we resum this as
$$\partial_{\hat{v}} W= \sum_{m,k}m d_{k,m}log(1-q^k e^{m \hat{v}})$$
where $q=e^{-t}$ and $d_{m,k}$ are expected
to be integers which label `the number of primitive
discs' wrapping ${\bf P}^1$ $k$ times and wrapping around
the $S^1$ of the Lagrangian submanifold $m$ times.
It is quite remarkable that, indeed, doing the resummation
we find that $d_{m,k}$ are integers, 
as far as we have checked (see table 1). Moreover
there are infinitely many non-vanishing integers, (unlike
the previous case where there were only two non-trivial
integers).  It would be quite interesting to
verify these numbers directly. Note also that the growth
of these numbers is as large as that observed for
the primitive rational curves:
for discs wrapping ${\bf P}^1$ a fixed number $k$
times and for large wrapping number $m$ on the $S^1$,
the degeneracies grow like $d_{k,m}\sim m^{2k-2}/{(k!)^2}$.

$$
\vbox{\offinterlineskip\tabskip=0pt \halign{\strut \vrule#&
{}~\hfil$#$~& \vrule#& {}~\hfil$#$~& {}~\hfil$#$~& {}~\hfil$#$~&
{}~\hfil$#$~& {}~\hfil$#$~& {}~\hfil$#$~& {}~\hfil$#$~&
{}~\hfil$#$~& \vrule$#$\cr \noalign{\hrule}
& m &&d_{0,m}&d_{1,m}&d_{2,m} &d_{3,m}
&d_{4,m}&d_{5,m}&d_{6,m}&\ldots & \cr
\noalign{\hrule} &1 &&1 &-1&     0   &        0&           0&
0&         0&&\cr &2 && 0 &-1&    1   &        0&           0&
0&         0&&\cr &3 && 0 &-1&    2   &        -1&           0&
0&         0&&\cr &4 && 0 &-1&    4   &        -5&          2&
0&         0&&\cr &5 && 0 &-1&    6   &       -14&         14&
-5&         0&&\cr &6 && 0 &-1&    9   &       -31&         52&
-42&       13&&\cr &7 && 0 &-1&   12   &       -60&        150&
-198&      132&&\cr &8 && 0 &-1&   16   &      -105&        360&
-693&      752&&\cr &9 && 0 &-1&   20   &      -171&        770&
-2002&     3114&&\cr &10&& 0 &-1&   25   &      -256&       1500&
-5045&    10514&&\cr &11&& 0 &-1&   30   &      -390&       2730&
-11466&    30578&&\cr &12&& 0 &-1&   36   &      -556&       4690&
-24024&    79420&&\cr &13&& 0 &-1&   42   &      -770&       7700&
-47124&   188496&&\cr &14&& 0 &-1&   49   &     -1040&      12152&
-87516&   415716&&\cr &15&& 0 &-1&   56   &     -1375&      18564&
-155195&  862194&&\cr &16&& 0 &-1&   64   &     -1785&
27552& -264537& 1697472&&\cr &\vdots&& & &       &         &
&        &         &&\cr \noalign{\hrule} } \hrule}$$
\noindent{\ninepoint\sl \baselineskip=8pt {\bf Table.1}: {\rm 
Holomorphic disc numbers for A-brane on $O(-1)+O(-1)$
over ${\bf P}^1$ in phase II.}}
\bigskip

\subsec{Degeneration of ${\bf P}^1\times {\bf P}^1$}
The computation of the superpotential in this case can be done
from the general formalism we have discussed for the Lagrangian
submanifolds ending on the toric skeleton.  However, to
check integrality properties one has to take into account the 
closed string mirror map since the quantum corrected areas $T_1,T_2$
are non-trivial functions of $t_1,t_2$. One should, thus, also
expect non-trivial analog of mirror map for the boundary
variables $u$, and $v$. To study this we consider a particular
limit of ${\bf P}^1\times {\bf P}^1$ where there
already is a non-trivial, but relatively simple mirror map.
This is the degenerate limit of ${\bf P}^1\times {\bf P}^1$ where the
size $t_2$ of the second ${\bf P}^1$ is taken to infinity. In this
limit, the equation of the mirror becomes:
\eqn\Pb{e^{u}+e^{-t_1-u}+e^{v}+1=0.}
There are, again, two phases (see figure 6):
\bigskip
\centerline{\epsfxsize 5.5 truein\epsfbox{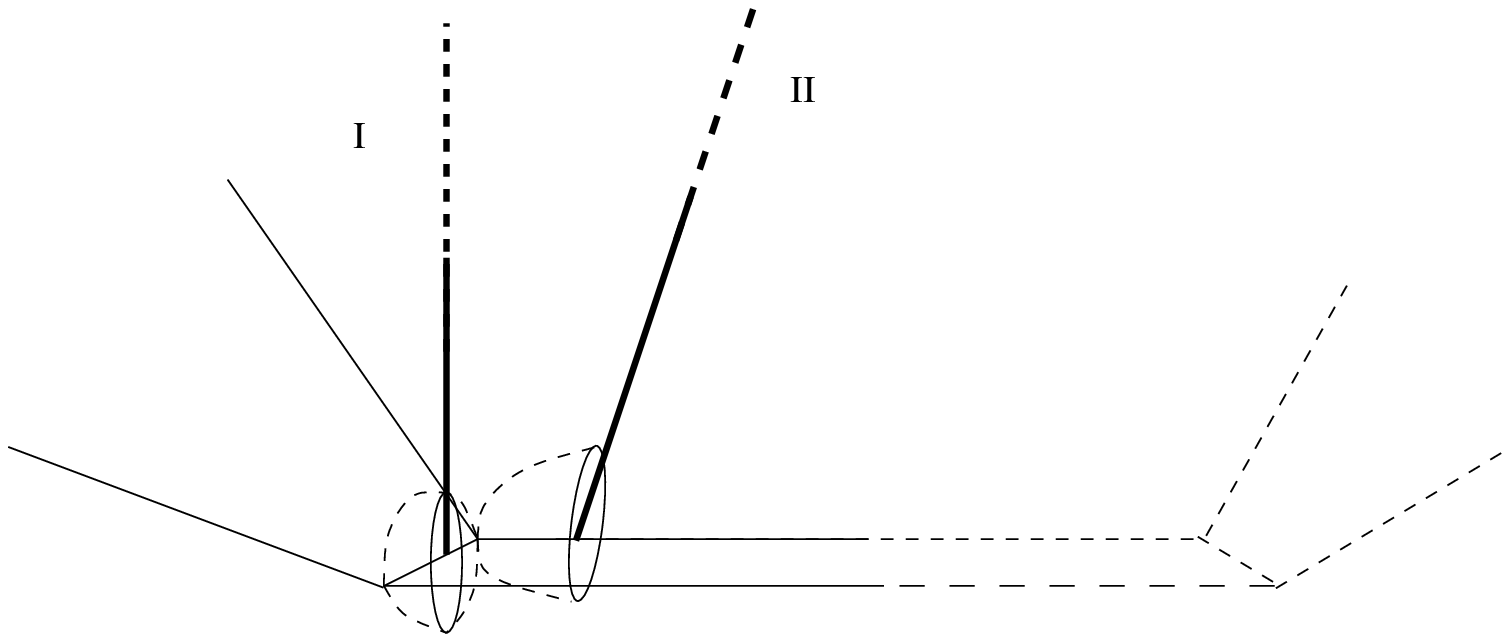}}
\rightskip 2pc
\noindent{\ninepoint\sl \baselineskip=8pt {\bf Fig.6}: {\rm 
Degenerate limit of ${\bf P}^1\times {\bf P}^1$ where the size of second 
${\bf P}^1$ is infinite.}}
\bigskip

\noindent{\ninepoint\sl \baselineskip=8pt {{\cal{Phase}} I}}

This is the phase in which the good variable on the curve 
is $u$ and 
$$v(u)=i\pi + log[e^{u}+e^{-t_1-u}+1].$$
In this case the mirror map gives $T_1$ in terms of $t_1$
\eqn\q{e^{-t_1} =\frac{q}{(1+q)^2},}
where $q=e^{-T_1}$ \ref\moras{P.~S.~Aspinwall, B.~R.~Greene and 
D.~R.~Morrison,``Measuring small distances in N=2 sigma models,''
Nucl.\ Phys.\  {\bf B420}, 184 (1994)[hep-th/9311042].}.  
It is natural to modify the boundary fields to ${\hat u}, {\hat v}$ such that
$$e^{\hat{u}}= -(1+q)e^u ~, \qquad e^{\hat{v}}= -(1+q)e^v.$$
To motivate this, note that 
$$e^{\hat{u}} e^{T_1/2}=-e^u e^{t_1/2}$$
which is consistent with the fact that when the Lagrangian
submanifold intersects the equator of the $\bf{P}^1$ both left hand side and
the right hand side of the equation should be one.
Using this we get

\eqn\potpA{\partial_{\hat{u}} W= -\hat{v}=
 -log[(1-e^{\hat{u}})(1-qe^{-\hat{u}})].}
Thus, there are again only two primitive disc instantons associated to two
hemispheres of the finite size ${\bf P}^1$, as expected (see figure 6).

\noindent{\ninepoint\sl \baselineskip=8pt {{\cal{Phase}} II}}

Solving for $\hat{u}$ in terms of $\hat{v}$ we find
\eqn\potpB{\partial_{\hat{v}} W= \hat{u}=log(\frac{1+q-e^{\hat{v}}}{2}+
\frac{\sqrt{(1+q-e^{\hat{v}})^2-4q}}{2}).}
Expanding this around the large radius limit we find
\eqn\potpb{\partial_{\hat{v}} W= log(1-e^{\hat{v}})+
\sum_{k=1}^{\infty}\sum_{m=1}^{\infty} C_{k,m}q^k e^{m\hat{v}}}
where,
\eqn\coeff{C_{k,m} =  -(-1)^k \frac{B(m,k)}{m+k} -
\sum_{n=1}^{k}(-1)^{k+n}\frac{B(n,n,m,k-n)}{k+n+m}.}
Again we find, remarkably, that the $d_{k,m}$ are integers (see
table 2),
and that the degeneracies of primitive discs grow 
like $d_{k,m}\sim m^{2k-1}$ for $k$ fixed and $m$ large.

It would be interesting to extend the mirror map computation
to the case where both $t_1,t_2$ are finite.  We are currently
investigating this case.
$$
\vbox{\offinterlineskip\tabskip=0.5pt \halign{\strut \vrule#&
{}~\hfil$#$~& \vrule#& {}~\hfil$#$~& {}~\hfil$#$~& {}~\hfil$#$~&
{}~\hfil$#$~& {}~\hfil$#$~& {}~\hfil$#$~& {}~\hfil$#$~&
{}~\hfil$#$~& \vrule$#$\cr \noalign{\hrule}
& m &&d_{0,m}&d_{1,m}&d_{2,m} &d_{3,m}
&d_{4,m}&d_{5,m}&d_{6,m}&\ldots & \cr
\noalign{\hrule} &1 && 1 &1&    1   &        1&           1& 1&
1&&\cr &2 && 0 &1&    2   &        4&           6& 9& 12&&\cr &3
&& 0 &1&    4   &       11&          25& 49& 87&&\cr &4 && 0 &1& 6
&       25&          76& 196& 440&&\cr &5 && 0 &1&    9   & 49&
196& 635& 1764&&\cr &6 && 0 &1&   12   &       87& 440& 1764&
5926&&\cr &7 && 0 &1&   16   &      144&         900& 4356&
17424&&\cr &8 && 0 &1&   20   &      225&        1700& 9801&
46004&&\cr &9 && 0 &1&   25   &      336&        3025& 20449&
111333&&\cr &10&& 0 &1&   30   &      484&        5110& 40080&
250488&&\cr &11&& 0 &1&   36   &      676&        8281& 74529&
529984&&\cr &12&& 0 &1&   42   &      920&       12936& 132496&
1063626&&\cr &13&& 0 &1&   49   &     1225& 19600& 226576&
2039184&&\cr &14&& 0 &1&   56   &     1600& 28896& 374544&
3755808&&\cr &15&& 0 &1&   64   &     2055& 41616& 600935&
6677055&&\cr &16&& 0 &1&   72   &     2601& 58680& 938961&
11502216&&\cr &\vdots&& & &       &         & & & &&\cr
\noalign{\hrule} } \hrule}$$
\noindent{\ninepoint\sl \baselineskip=0.5pt {\bf Table.2}: {\rm 
Holomorphic disc instanton numbers for degeneration of
${\bf P}^1\times {\bf P}^1 $.}}

\centerline{\bf Acknowledgements}
We would like to thank K. Hori, S. Katz and R. Thomas
for valuable discussions.
This research is supported in part by NSF grants PHY-9802709
and DMS 9709694.

\listrefs
\end